\documentclass[superscriptaddress,aps,preprintnumbers,amsmath,showpacs,amssymb,prd,nofootinbib,reprint,floatfix]{revtex4-1}
\usepackage{bm, color} 
 \usepackage{amssymb,amsfonts,slashed,amsthm,amsmath,graphicx, soul}

\usepackage{footmisc}
\usepackage{hyperref}
\usepackage{color}
\usepackage[normalem]{ulem}

\begin{document}

\renewcommand{\thesection}{\Alph{section}}

\newcommand{\mplank}{\textrm{M}_{\textrm{P}}}
\newcommand{\mg}{m_{\gamma^{\prime}}}
\newcommand{\higgs}{H_{\scriptsize \rm h}}
\newcommand{\higgst}{\tilde{H}_{\scriptsize \rm h}}
\renewcommand{\Re}{\mathrm{Re}}
\newcommand{\MF}{{\sf B}}

\newcommand{\muu}{m_{\gamma^{\prime}}}
\newcommand{\chie}{\chi_{\rm eff}}
\newcommand{\ve}[2]{\left(
\begin{array}{c}
 #1\\
#2
\end{array}
\right)
}

\renewcommand\({\left(}
\renewcommand\){\right)}
\renewcommand\[{\left[}
\renewcommand\]{\right]}

\def\ebq{\end{equation} \begin{equation}}
\renewcommand{\figurename}{Figure.}
\renewcommand{\tablename}{Table.}
\newcommand{\Slash}[1]{{\ooalign{\hfil#1\hfil\crcr\raise.167ex\hbox{/}}}}
\newcommand{\bra}[1]{ \langle {#1} | }
\newcommand{\ket}[1]{ | {#1} \rangle }
\newcommand{\beq}{\begin{equation}}  \newcommand{\eeq}{\end{equation}}
\newcommand{\bef}{\begin{figure}}  \newcommand{\eef}{\end{figure}}
\newcommand{\bec}{\begin{center}}  \newcommand{\eec}{\end{center}}
\newcommand{\non}{\nonumber}  \newcommand{\eqn}[1]{\begin{equation} {#1}\end{equation}}
\newcommand{\laq}[1]{\label{eq:#1}}  
\newcommand{\dd}[1]{{d \o d{#1}}}
\newcommand{\Eq}[1]{Eq.~(\ref{eq:#1})}
\newcommand{\Eqs}[1]{Eqs.~(\ref{eq:#1})}
\newcommand{\eq}[1]{(\ref{eq:#1})}
\newcommand{\Sec}[1]{Sec.\ref{chap:#1}}
\newcommand{\ab}[1]{\left|{#1}\right|}
\newcommand{\vev}[1]{ \left\langle {#1} \right\rangle }
\newcommand{\bs}[1]{ {\boldsymbol {#1}} }
\newcommand{\lac}[1]{\label{chap:#1}}
\newcommand{\SU}[1]{{\rm SU{#1} } }
\newcommand{\SO}[1]{{\rm SO{#1}} }

\def\({\left(}
\def\){\right)}
\def\dt{{d \o dt}}
\def\diag{\mathop{\rm diag}\nolimits}
\def\Spin{\mathop{\rm Spin}}
\def\O{\mathcal{O}}
\def\U{\mathop{\rm U}}
\def\Sp{\mathop{\rm Sp}}
\def\SL{\mathop{\rm SL}}
\def\tr{\mathop{\rm tr}}
\newcommand{\OR}{~{\rm or}~}
\newcommand{\AND}{~{\rm and}~}
\newcommand{\EV}{ {\rm \, eV} }
\newcommand{\KEV}{ {\rm \, keV} }
\newcommand{\MEV}{ {\rm \, MeV} }
\newcommand{\GEV}{ {\rm \, GeV} }
\newcommand{\TEV}{ {\rm \, TeV} }

\def\o{\over}
\def\a{\alpha}
\def\b{\beta}
\def\c{\varepsilon}
\def\d{\delta}
\def\e{\epsilon}
\def\f{\phi}
\def\g{\gamma}
\def\h{\theta}
\def\k{\kappa}
\def\l{\lambda}
\def\m{\mu}
\def\n{\nu}
\def\p{\psi}
\def\q{\partial}
\def\r{\rho}
\def\s{\sigma}
\def\t{\tau}
\def\u{\upsilon}
\def\v{\varphi}
\def\w{\omega}
\def\x{\xi}
\def\y{\eta}
\def\z{\zeta}
\def\D{\Delta}
\def\G{\Gamma}
\def\H{\Theta}
\def\L{\Lambda}
\def\F{\Phi}
\def\P{\Psi}
\def\S{\Sigma}
\def\me{\mathrm e}
\def\ol{\overline}
\def\tl{\tilde}
\def\*{\dagger}

\newcommand{\exclude}[1]{}

\def\bra{\langle}
\def\ket{\rangle}
\def\beq{\begin{equation}}
\def\eeq{\end{equation}}
\newcommand{\C}[1]{\mathcal{#1}}
\def\ov{\overline}

\hbadness=20000

\preprint{TU-1134}
\title{Shining ALP Dark Radiation}

\author{Joerg Jaeckel}
\affiliation{Institut f\"ur theoretische Physik, Universit\"at Heidelberg,
 Philosophenweg 16, 69120 Heidelberg, Germany}
\author{Wen Yin}
\affiliation{Department of Physics, Tohoku University, Sendai, Miyagi 980-8578, Japan }

\begin{abstract}
String scenarios typically not only predict axion-like particles (ALPs) but also 
significant amounts of ALP dark radiation originating from the decay of the inflaton or a more general modulus.
In this paper, we study  the decay of such non-thermally produced relativistic (but massive) ALPs to photons.
If the ALPs are sufficiently highly energetic, contribute to $\D N_{\rm eff} \gtrsim \O(0.001)$ and {have} a mass $m_a\gtrsim \MEV$ we find that, using observations of X-, and $\g$-rays, the CMB and BBN, very small values of the ALP-photon coupling can be probed, corresponding to an origin of this coupling at the string (or even Planck) scale.
\noindent
\end{abstract}
\maketitle
\flushbottom

\section{Introduction}

In string or M-theory, ``axions'' are ubiquitous and one of them may even be the QCD axion~\cite{Witten:1984dg, Svrcek:2006yi,Conlon:2006tq,Arvanitaki:2009fg,Acharya:2010zx, Higaki:2011me, Cicoli:2012sz,Demirtas:2018akl,Mehta:2020kwu,Mehta:2021pwf}. In particular, it is likely that one or more of these axions couples to the standard model (SM) photon in M-theory~\cite{Acharya:2010zx}.
In the following we will focus on such ``axions'' coupled to two-photons but heavier than the QCD axion. To make this distinction explicit we will refer to them as axion-like particles (ALPs). (For some reviews of axions or ALPs see~\cite{Jaeckel:2010ni,Ringwald:2012hr,Arias:2012az,Graham:2015ouw,Marsh:2015xka,Irastorza:2018dyq,DiLuzio:2020wdo}.)
The masses of such ALPs are generated by non-perturbative effects and therefore are expected to feature a logarithmic scaling~\cite{Svrcek:2006yi,Conlon:2006tq,Arvanitaki:2009fg,Acharya:2010zx, Cicoli:2012sz,Gherghetta:2020keg,Kitano:2021fdl}.\footnote{In Ref.\cite{Marsh:2019bjr}, it was mentioned that the ALP mass in the M-theory~\cite{Acharya:2010zx} is sensitive to the running of the gauge couplings. Here we note that the axion may be heavy in the gauge mediation scenario of  supersymmetry breaking, in which case the gauge coupling, $\alpha_{\rm GUT},$ is much stronger than $1/25$ at the string scale and the ALP mass $\propto \exp{[-2\pi/\a_{\rm GUT}]}$ is  significantly enhanced.} 
They can be either heavy or light but are unlikely to be massless since quantum gravity is argued to break any global symmetry, e.g.~\cite{Misner:1957mt,Banks:1988yz, Barr:1992qq,Kamionkowski:1992mf,Holman:1992us,Kallosh:1995hi,Banks:2010zn,Harlow:2018tng,Alvey:2020nyh, Yonekura:2020ino}\footnote{Interestingly a non-vanishing fraction of the early works~\cite{Kamionkowski:1992mf,Holman:1992us,Kallosh:1995hi} discusses axions.} 
(for a nice introduction of the string landscape see, e.g.~\cite{Hebecker:2020aqr}).

In the early Universe, it is plausible, perhaps even likely that reheating proceeds via the decay of a modulus, which can also be the inflaton itself. The super partner of such a modulus may be the axion-like particle that we are interested in. Then, if kinematically allowed, the ALPs are naturally produced in the modulus' decays.  
Those ALPs contribute to the dark radiation.  
Such dark radiation ALPs are widely discussed as a Cosmic axion Background (CaB)~\cite{ Cicoli:2012aq,Higaki:2012ar,Angus:2013sua, Conlon:2013isa, Hebecker:2014gka, Evoli:2016zhj, Armengaud:2019uso, Dror:2021nyr,Jaeckel:2021gah}. Alternatively or even additionally ALP dark radiation may arise from a relatively long-lived sub-dominant species that has a sizable branching fraction into ALPs.
Tests of the CaB using ALP-photon conversion in Earth based experiments were studied in~\cite{Conlon:2013isa, Armengaud:2019uso, Dror:2021nyr}. 
In addition ALP-photon conversion may  occur in astrophysical and cosmological magnetic fields, leaving potentially detectable imprints~\cite{Higaki:2013qka,Fairbairn:2013gsa,Conlon:2013txa, Tashiro:2013yea, Payez:2014xsa, Marsh:2017yvc, Reynolds:2019uqt}.

ALPs may also be produced via thermal scattering if they couple to the standard model particles. This  yields very strong cosmological  bounds~\cite{Turner:1986tb,Chang:1993gm, Moroi:1998qs,Hannestad:2005df, Cadamuro:2011ti,Cadamuro:2011fd,Jaeckel:2014qea,Salvio:2013iaa,Daido:2017tbr}.  
However, those are typically contingent on a sufficiently high reheating temperature or large ALP photon coupling such that the ALPs thermalize~\cite{Cadamuro:2011fd}. In contrast, the situation outlined above and focused on in the present paper is usually associated with rather low reheating temperatures or small ALP photon coupling such that the thermal bounds will often not apply.

In this paper, we carefully study
a situation where the ALPs constituting ALP dark radiation, produced from the decay of a precursor particle, themselves decay to photons\footnote{There are also several studies discussing the decays of heavy ALPs emitted from astrophysical objects, in particular supernovae, around the present Universe~\cite{Giannotti:2010ty, Jaeckel:2017tud, Calore:2021klc}. }.
We estimate carefully the  dark radiation contribution to the deviation of the effective number of neutrino, $\D N_{\rm eff}$, by taking  account of the decoupling effect. If this is measured in the future by cosmic microwave background (CMB) and baryonic acoustic oscillation (BAO) experiments~\cite{Kogut:2011xw, Abazajian:2016yjj, Baumann:2017lmt}, it is a probe of the ALP radiation at the recombination epoch.
Then, we study the ALP decays to photons by taking account of their mass.
This requires sufficiently massive ALPs and therefore complements the tests via conversion which usually focus on sub-eV masses.
For the mother particle decay we consider both cases that the precursor particle decays before and after reheating, the latter of which includes the present period meaning that the mother particle is part of the dark matter (DM).  
We also discuss the possibility that the ALP radiation becomes non-relativistic after the recombination.
We find that such decaying ALPs can be constrained and tested up to very large decay constants, even up to the string scale if the mass $m_a \gtrsim 1\MEV.$
In this way string or M-  theory may be tested via future X- and $\gamma$-ray observations by e.g. ATHENA~\cite{Barret:2018qft}, CTA~\cite{CTAConsortium:2010umy},  
eROSITA~\cite{eROSITA:2012lfj}, Fermi-Lat~\cite{Fermi-LAT:2012edv,Fermiweb}, GAMMA-400~\cite{Galper:2012fp},  XRISM~\cite{XRISMScienceTeam:2020rvx}, and the CMB spectra, in addition to the $\D N_{\rm eff}$ measurement.

\vspace{0.2cm}

\section{Review of ALP radiation from the decay of heavy particles}

Let us briefly recall the main features of the scenario we want to consider, mainly following~\cite{Jaeckel:2021gah} but similar setups can also be found in~\cite{Hebecker:2014gka, Evoli:2016zhj, Armengaud:2019uso, Dror:2021nyr}.

For concreteness, let us consider a modulus field, $\f$.  
Our discussion does not change if this modulus is replaced by 
any other heavy particle, such as an inflaton or a heavy gravitino, that has two-body decays into an ALP in the early stages of the Universe. 
In the following, unless otherwise stated and especially for the numerical estimates, 
we assume that $\phi$ at one point dominated the Universe and decays to reheat the Universe.
Let us stress, however, that even if the mother particle does not dominate the Universe, the bounds we will derive do not change if we replace the reheating temperature, $T_R$, by the decay temperature, $T_\f$, with the same definition, given below in \Eq{reh}.
The mother particle may be produced thermally,  non-thermally including gravitationally, by misalignment~\cite{Preskill:1982cy,Abbott:1982af,Dine:1982ah} or via other particles decay etc.. The only feature we make use of is that it is non-relativistic at the time of decay. But even if this is not the case we expect that qualitatively similar results can be obtained in many other situations.

The total $\f$ decay rate to the standard model particles via higher dimensional terms can be given in the form of 
\beq
\label{eq:kappadef}
\G_\f = \k \frac{m_\f^3}{ M^2}.
\eeq 
Here $m_\f$ is the modulus mass, and $\k$ is a coefficient encoding the details of the decay. $M$ is the scale of the higher dimensional term giving rise to the interaction, and it may be regarded as the cutoff scale of the theory.
Allowing for two- or three-body phase space suppression we expect $\k \sim \O(0.1)- \O(10^{-3})$. 
Typical moduli but also the gravitino usually behave in this manner with $M\sim M_{\rm pl}$. 

Smaller values of $\kappa$ may be motivated by requiring that the radiative corrections to the mass squared caused by the higher dimensional interaction are smaller than $m_\f^2$, giving, $\k \lesssim m_\f^2/M^2.$  
{This is the} case if the mother particle is another ALP, which is stabilized in a CP-violating vacuum, decaying into the ALP or SM particles via mixing with daughter ALP or heavy modulus. Then the decay rate is suppressed by the mixing (see e.g. the Appendix A of \cite{Kim:2021eye}).
Moreover, requiring that the coupling does not cause a too large (tree or radiative) correction to the Higgs mass squared $\k \lesssim (100\GEV/M)^2 $ may be needed for large $m_\f$ (see~\cite{Jaeckel:2020oet} for a more detailed discussion). 
This may be particularly important if the supersymmetry scale is as high as $M$. 
Those radiative correction arguments are usually used in the context that $\f$ is the inflaton which needs a flat potential to satisfy the slow-roll conditions. Therefore, $\k$ is model dependent.

The decay reheats the Universe with the reheating temperature defined by $T_R\equiv (g_{\star} \pi^2/90)^{-1/4}\sqrt{\G_\f M_{\rm pl}}$. This yields,
\beq
\laq{reh}
T_R \approx 20 \MEV \sqrt{\k  } \(\frac{M_{\rm pl}}{M}\) \(\frac{11}{g_{\star}}\)^{1/4}\( \frac{m_\f}{100\TEV}\)^{3/2},
\eeq
where we use $M_{\rm pl}\approx 2.4\times 10^{18}\GEV$ for the reduced Planck mass.
As indicated by the benchmark values in the equation, such scenarios often have a quite low reheating temperature.
The decay temperature, $T_\phi$, when $\f$ does not dominate the Universe can be obtained analogously from $H=\Gamma_\phi$, where $H$ is the Hubble parameter at the decay evaluated in a standard $\L$CDM scenario.

Below we will recall how to calculate the spectrum of the ALPs resulting from the $\f$ decay in the case where it is dominating the energy density.
This requires the density of $\f$ as an input but the initial density of radiation is not important. As described in~\cite{Jaeckel:2021gah} this can be obtained by solving the coupled Boltzmann equations for the density $\rho_\f$ and the entropy density $s_r$, completed by expressions for the Hubble rate $H$, the temperature $T$ and the radiation energy density $\rho_r$,
\begin{align}
\laq{1}
&\dot{\rho}_\f+3H\rho_\f  = -\G_{\f} \rho_\f \,, \\
&\laq{2}\dot{s}_r+3H s_r = c[t] \G_{\f} \rho_\f \,,\\
& c[t] = \frac{4\left(T g_{s\star}'+3 g_{s\star}\right)}{3 T \left(T g'_\star+4 g_\star\right)}\,,\\[0.2cm]
&H\approx \sqrt{\frac{\rho_\f+\rho_r}{3M_{\rm pl}^2}}\,,\\
&T=\left(\frac{45}{2\pi^2 g_{s\star}} s_{r}\right)^{1/3}\,,\\
&\rho_r=\frac{\pi^2 g_\star }{30} T^4\,.
\end{align}
The final inputs for our calculations are $g_\star \AND g_{s\star}$, the relativistic degrees of freedoms of the energy density and the entropy density from~\cite{Husdal:2016haj}.

\bigskip

The modulus naturally decays into ALPs that are kinematically available.
In particular, it can couple to ALPs, $a$, via e.g. 
\cite{Cicoli:2012aq,Higaki:2012ar,Angus:2013sua, Conlon:2013isa, Higaki:2013lra, Hebecker:2014gka,Moroi:2020has}
\beq
{\cal  L} \supset  \frac{\phi}{M_a} \partial_\mu a \partial^\mu a,
\eeq
where $1/M_a$ is the coupling between $\phi$ and $a$. 
A noteworthy case is that the modulus is the superpartner of the ALP, i.e. the saxion, or {if it} mixes with the saxion.
Via this interaction the decay 
\beq
\f \to a a 
\eeq
is possible.

The decay rate of $\f \to aa$ is given by \cite{Cicoli:2012aq,Higaki:2012ar,Angus:2013sua, Conlon:2013isa, Higaki:2013lra, Hebecker:2014gka,Moroi:2020has} 
\beq
  \Gamma_{\phi\rightarrow aa} =
  \frac{1}{32\pi} \frac{m_\phi^3}{M_a^2},
\eeq
for $m_\f\gg m_a$ with $m_a$ being the ALP mass.
A component of the energy is transferred into the ALP if the branching fraction 
\beq
B_{\f\to aa}\equiv \frac{ \Gamma_{\phi\rightarrow aa}}{\G_\f}
\eeq
is non-vanishing. 
Nevertheless, in the following we assume that $B_{\f \to aa} \ll 1$, i.e. ALP production is not dominant, so that we can neglect its component in the expansion history of the Universe and satisfy the constraint on the expansion history around the BBN and recombination  eras~\cite{Aghanim:2018eyx,Fields:2019pfx}. 
Note, however, that $B_{\f \to aa}\sim 1$ is possible when $\f$ does not dominate the Universe.

Such a component, if relativistic until the recombination epoch, contributes to the dark radiation as (cf., e.g., Refs.\,\cite{Choi:1996vz,Cicoli:2012aq,Higaki:2012ar, Conlon:2013isa,Jaeckel:2020oet}),
\beq
 \D N_{\rm eff}
 \sim 6.1 B_{\f\to aa}  \(\frac{11}{{g_{s\star }(T_R)}^{4}g_\star(T_R)^{-3}} \)^{1/3}\,. \laq{Neff}
\eeq
This can be derived analytically by assuming constant $g_{\star},\,  g_{s\star}$ during the $\f$ decay.
Note that, as long as the ALPs are sufficiently relativistic $\D N_{\rm eff}$ does not depend on the mass or energy of the ALP. A more precise result, obtained by solving the relevant Boltzmann equations and integrating over the resulting ALP spectrum is shown in Fig.~\ref{fig:Neff}. This demonstrates the good accuracy of the formula given above. That said we have to be mindful of the effect of the decrease of $g_{\star}, g_{s \star}$
if we perform a precise analysis. 
The coefficient of \Eq{Neff}, which is usually used in the literature, is not clearly defined when $g_{\star},\,  g_{s\star}$ change rapidly. This is obvious because it is then unclear which definition of $g_{\star},\,  g_{s\star}$ we should use in \Eq{Neff},
despite $\Delta N_{\rm eff}$ (red points) being a smooth function. 
In fact, the entropy release can enhance $\D N_{\rm eff}$ above the naive estimate obtained by assuming a constant $g_{\star},\,  g_{s\star}$. This is because the thermalized radiation energy density scales as $\rho_r \propto s_r^{4/3} g_{s\star}^{-4/3} g_{\star}$. $\D N_{\rm eff} \propto \rho_a s_r^{-4/3} \sim Br_{\phi \to aa} \rho_r s_r^{-4/3}\propto g_{s\star}^{-4/3} g_{\star}$. We have used $\rho_a \sim Br_{\phi\to aa} \rho_{r}$, which is justified in such a short period that the expansion of Universe can be neglected. 
When $g_{s\star}\simeq g_{\star}$ decreases during the short period for $\phi$ decays, $\D N_{\rm eff}$ increases. This is also found in the figure by comparing the numerical and analytical results. In the case that $\f$ does not dominate the Universe and decays much after BBN, the analytic estimation is good enough.

It is noteworthy that ALPs contributing to $\D N_{\rm eff}>\O(0.01)$ at around the recombination era can be measured in future CMB and BAO experiments \cite{Kogut:2011xw, Abazajian:2016yjj, Baumann:2017lmt}.

\begin{figure}[t!]
\begin{center}  
\includegraphics[width=75mm]{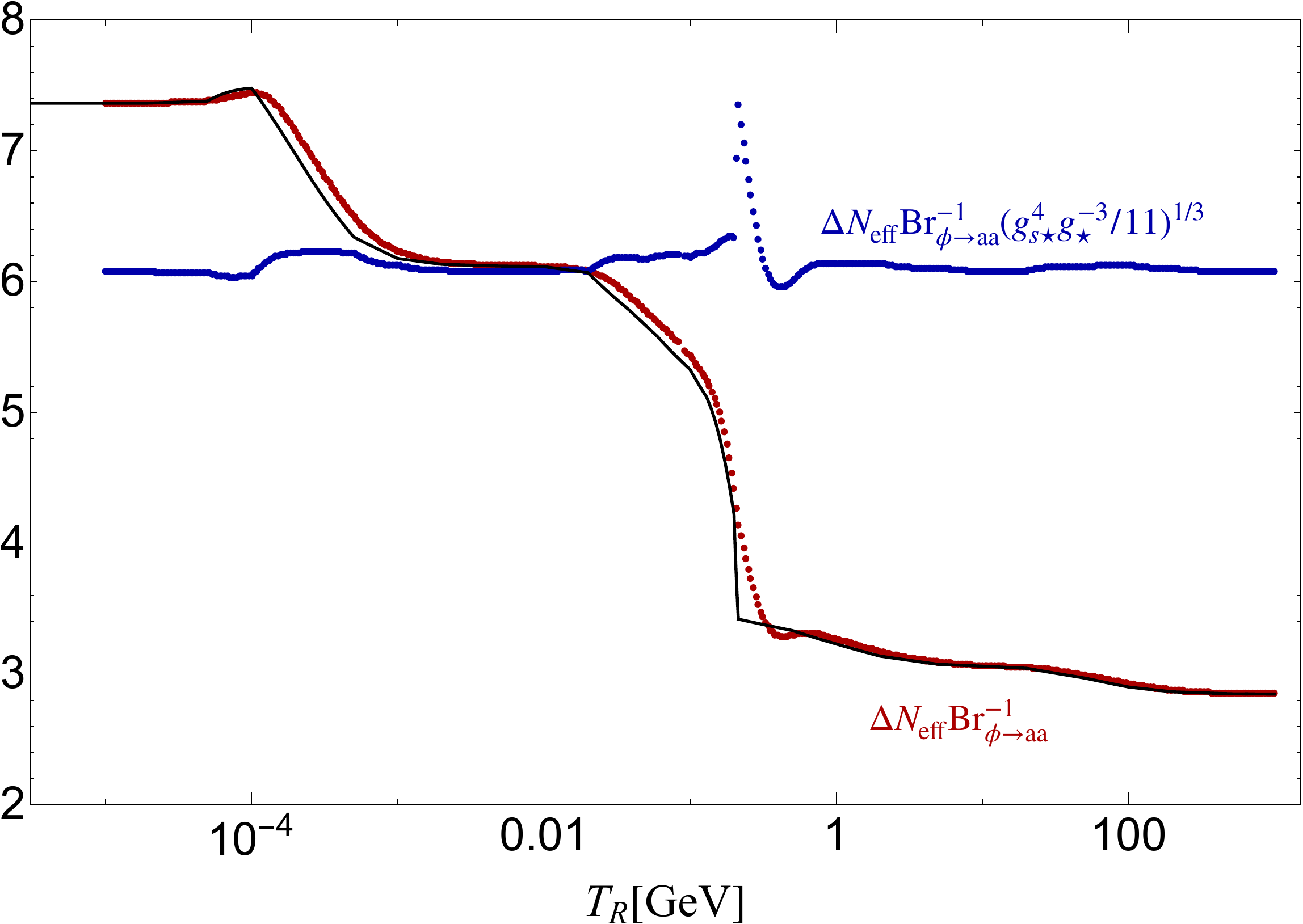}
\end{center}
\caption{Analytical (black solid line) and numerical (red points) results for $\D N_{\rm eff} B_{\phi\to aa}^{-1}$ 
by varying $T_R$ and assuming that the relativistic $a$ does not decay until the present. 
We also show  
$ \D N_{\rm eff} B_{\phi\to aa}^{-1}  \(\frac{11}{{g_{\star s}(T_R)}^{4}g_\star(T_R)^{-3}} \)^{-1/3}$ (blue points), for easy comparison with \Eq{Neff}.
}
\label{fig:Neff}
\end{figure}

\bigskip

The log-differential energy density of $a$ with momentum $p_a$  at $t$ is given by (see, again,~\cite{Jaeckel:2021gah} for details but note that here we use a log-energy differential distribution),
\begin{align}
\laq{rhoap}
\rho_{a, p_a}[t]&\equiv \frac{p_a^3 \sqrt{p_a^2+m_a^2}}{2\pi^2}f_{a, p_a}[t] \\
\laq{approx}&= 16E_a^4  \frac{\G_{\f \to aa}\rho_{\f}(t')}{H(t') m_\f^4}\theta(t-t'),
\end{align}
where $f_{a,k}$ is the distribution function of $a$, $E_a=\sqrt{p_a^2+m_a^2}$ is the energy of $a$, and we have neglected $m_a$ in the last line. 
Note that, this log-differential distribution is defined such that the total ALP energy density is given by $\rho_a= \int \rho_{a, p_a} d\log p_a.$
We will assume that the ALPs remain relativistic during the time-scales of our interest. 
Moreover, $t'$ is related with $t$ by $R[t']m_\f/2=R[t] p_a$ (cf.~\cite{Jaeckel:2021gah}), where $R$ denotes the scale factor related to the redshift, $z$, via $1+z=1/R$.
We also define $\hat{p}_a$ by $p_a=(1+z)\hat{p}_a$ for $p_a$ at the redshift $z$. 
If it travels freely up to the present, the ALP today has a typical momentum of 
\beq
 p^{\rm peak}_a\simeq 10^{-13}\GEV \frac{m_\f}{ g_{\star s}[T_R]^{1/3}T_R}. 
\eeq
For $M=M_{\rm pl}$ and $m_{\f}=100\,{\rm TeV}$ and $\k$ (defined in Eq.~\eqref{eq:kappadef}) in the range $0.001-0.1$ this formula gives $p^{\rm peak}_a\sim 1-10{\rm keV}$. 

Higher peak momenta can be achieved by lowering the reheating temperature.
Note, however, for temperatures lower than about an MeV the modulus cannot dominate the Universe in order to have successful BBN. 
One can also allow for larger $m_{\f}$. To realize this, while keeping $T_R$ fixed, one needs to have very small values of $\kappa$ (i.e. the modulus need to be in some sense sequestered from the SM and the axion and have no other decay channels; see also the discussion above).
More generally, we can consider the case where the abundance of the original modulus is suppressed, but potentially with a larger branching fraction to axions. The reheating temperature $T_R$ is then replaced by the decay temperature $T_{\f}$. In this case larger values of the peak momenta are possible without violating the observational constraints from reheating.

To illustrate the range of achievable values we show $p^{\rm peak}_a$ by varying the decay temperature, $T_\phi$ (which is obtained from the equation $H=\Gamma_\phi$ where we neglect the axion or modulus contribution in $H$),  or  the reheating temperature, $T_R$, in Fig.~\ref{fig:pa}. We find that if the modulus decays after the recombination temperature $\sim \EV$ we can have peak momenta above an MeV (blue lines). 
Moreover, if we consider a situation where radiative corrections need to be suppressed, e.g., if $\phi$ is the inflaton or if the supersymmetric scale is high, this will result {in} a larger $p_a^{\rm peak}$ for a given reheating/decay temperature (red lines). 
We also {indicate} the lower bound of the reheating temperature (purple band) which is needed for the success of big-bang nucleosynthesis~\cite{Kawasaki:1999na, Kawasaki:2000en,Hannestad:2004px, Ichikawa:2006vm, DeBernardis:2008zz, deSalas:2015glj, Kawasaki:2017bqm, Hufnagel:2018bjp, Forestell:2018txr, Hasegawa:2019jsa, Kawasaki:2020qxm,Depta:2020zbh}.  
Following this, we consider in the present paper rather large ranges of the ALP peak momentum. 
Results for a generic modulus of $\k=0.001-1$ that is reheating the Universe are shown in {Figs. \ref{fig:41} and \ref{fig:44}}. In addition, we in Figs.~\ref{fig:42} and \ref{fig:43} we also show the results with larger $p_{a}^{\rm peak}$ to take account of more general situations.

\begin{figure}[t!]
\begin{center}  
\includegraphics[width=85mm]{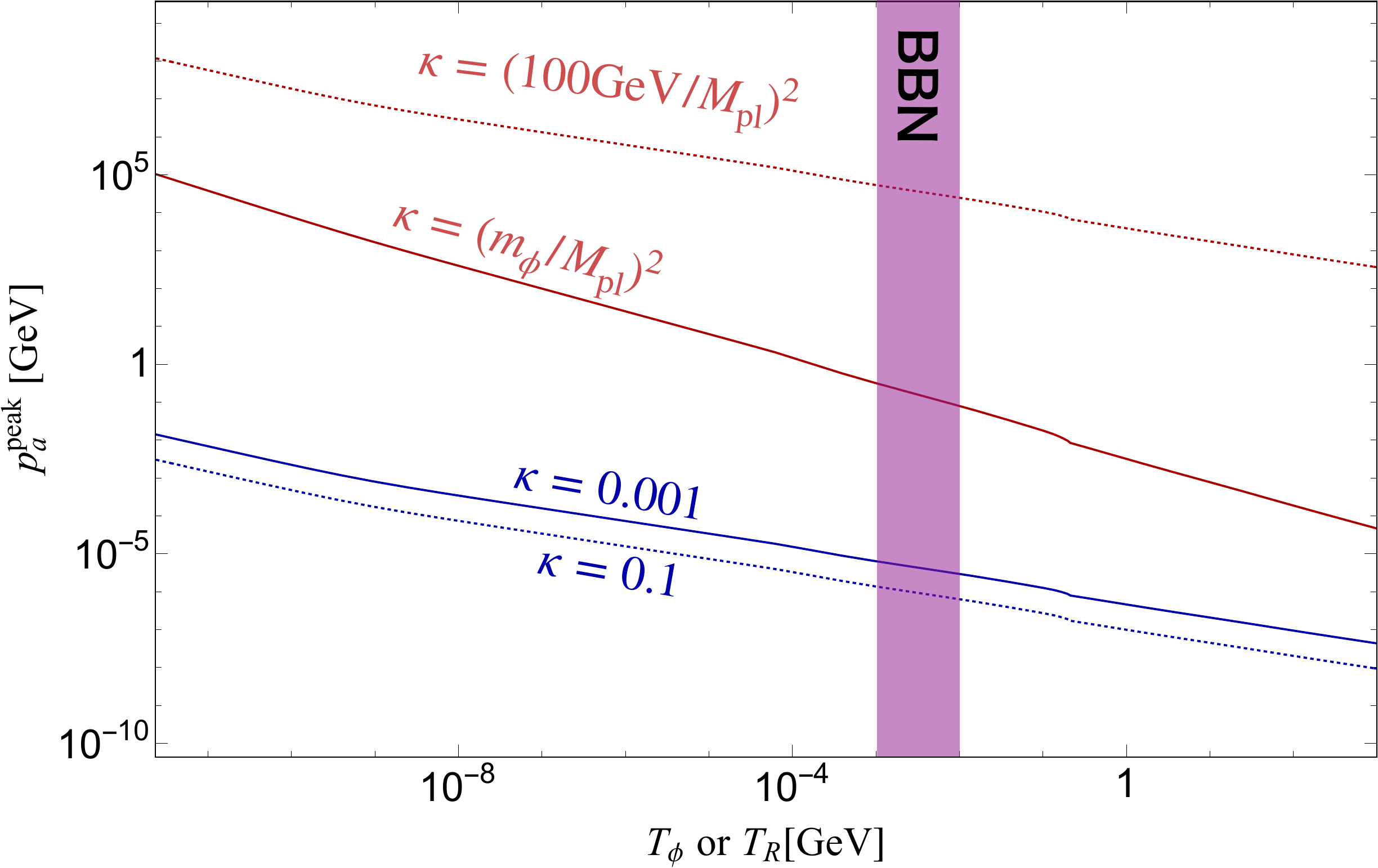}
\end{center}
\caption{Typical peak energy and decay/reheating temperature for different scenarios. Along the lines we vary $m_\phi$. Two fairly generic cases are $\kappa= 0.1$ (blue dashed line) and  $0.001$ (blue solid line) with $M=M_{\rm pl}$. We also show two cases where $\kappa$ is suppressed (red lines). In the case $m_\phi^2$ (weak scale$^2$) the respective scale is stable under radiative correction of $O(M_{\rm pl})^2$.
}
\label{fig:pa}
\end{figure}

As already mentioned some ALPs may couple to SM particles and, in particular, photons via 
\beq
{\cal L}\supset \frac{g_{a\g\g} }{4 }a  F_{\mu\n}\tl{F}^{\m\n},
\eeq
where $F_{\m\n}$  ($\tl{F}^{\m\n}$) is the field strength of the SM photon (its dual), and $g_{a \g\g} $ is the photon coupling. 
If the scattering rate between ALPs and plasma photons (or other weak bosons in the symmetric phase) is lower than the expansion rate, $C_{\rm th} g_{a \g\g}^2 T_R^3 \lesssim \sqrt{g_\star \pi^2/90 } T_R^2/M_{\rm pl}$, i.e. (cf. also~\cite{Cadamuro:2011fd})
\begin{equation}
\label{eq:equilibrium}
    g_{a \g\g}\lesssim 10^{-9}\GEV \frac{0.01}{C_{\rm th}} \(\frac{100\GEV}{T_R}\)^{1/2},
\end{equation} 
the ALP rarely interacts with the ambient plasma. 
Here, $C_{\rm th}$ is a model-coefficient depending on how $a$ couples to gauge fields in the symmetric phase.

If light, such dark radiation ALPs have the potential to be detected via ALP-photon conversion~\cite{Conlon:2013isa, Armengaud:2019uso, Dror:2021nyr,Higaki:2013qka,Fairbairn:2013gsa,Conlon:2013txa, Tashiro:2013yea, Payez:2014xsa, Marsh:2017yvc, Reynolds:2019uqt}.
However, in this paper, we will use the same coupling to study the decay of heavier ALPs. We will find that if $m_a>\O(\KEV) $ the constraints
can be stronger than any existing constraints. Future observations of such photons will provide a nice opportunity to search for the dark radiation and traces of reheating. 

Also note, that Eq.~\eqref{eq:equilibrium} delineates the opposite case of the thermal production that was used in~\cite{Cadamuro:2011ti,Cadamuro:2011fd}. In this sense the bound obtained therein and the ones we will derive complement each other.

\bigskip

\section{Decaying dark radiation}

Now let us focus on  the decay of the ALP radiation. 
Including the Lorentz factor $E_a/m_a$, the ALP decays into the photon pair with an energy dependent width,
\beq
\G_a[p_a]= \frac{g_{a\g\g}^2}{64\pi} \frac{m_a^4}{E_a}.
\eeq
When {$p_a \gg m_a$,} we have,
\begin{equation}
\G_a[p_a]\simeq \frac{g_{a\g\g}^2}{64\pi} \frac{m_a^4}{\hat{p}_a (1+z)},
\end{equation}
and for $p_a\ll m_a$, we have the usual
\beq 
\G_a[p_a]= \frac{g_{a\g\g}^2}{64\pi} m_a^3.
\eeq

 \begin{figure}[t!]
\begin{center}  
\includegraphics[width=85mm]{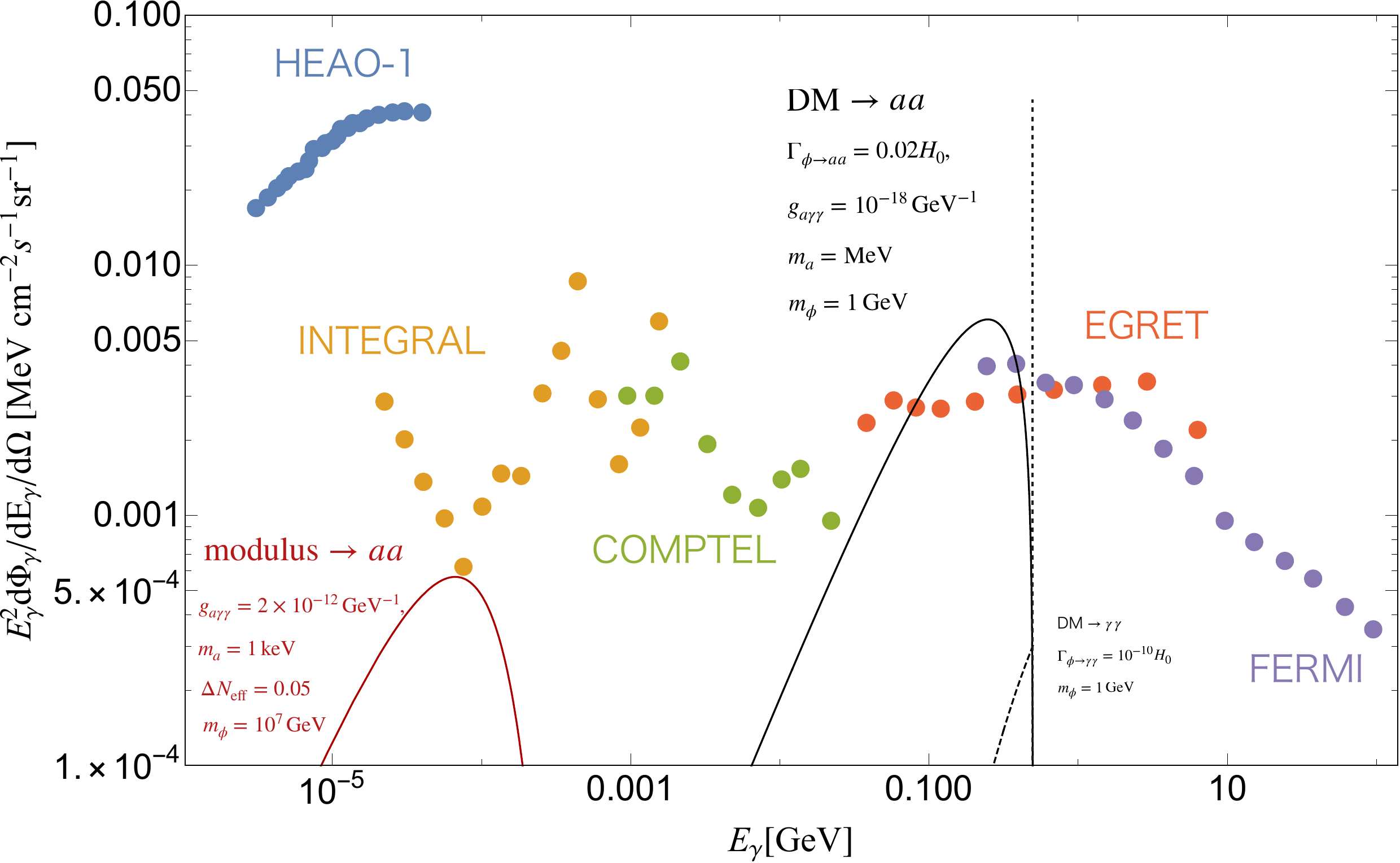}
\end{center}
\caption{The photon spectrum from the decay of ALP dark radiation {itself} originating from a modulus decay (red solid line) in the early Universe or from DM decay (black solid line). 
We take $g_{a\g\g}=2\times 10^{-12}\GEV^{-1}, m_a=1\KEV, m_\f=10^7\GEV, T_R=10\MEV, \AND \D N_{\rm eff}=0.05$ for case of relativistic ALPs from modulus decays. 
For the DM decay case we take $\Gamma_{\phi \to aa}=0.02H_0, g_{a\g\g}= 10^{-18}\GEV^{-1}, m_a=\MEV,$ with $H_0$ being the Hubble constant.  
For comparison we also show the ordinary photon spectrum from direct DM decay with $\Gamma_{\f\to aa}= 10^{-10}H_0 $ and DM mass $m_\f=1\,\GEV.$
The points indicate the photon flux observed from HEAO-1~\cite{Gruber:1999yr}, INTEGRAL~\cite{Bouchet:2008rp}, COMPTEL~\cite{kappadath1998measurement}, EGRET~\cite{Strong:2004de} and FERMI~\cite{Fermi-LAT:2012edv} experiments. 
 All of these are adopted from Ref.\,\cite{Essig:2013goa}. 
}
\label{fig:1}
\end{figure}

First, let us consider the decay of the ALP when it is relativistic. Then, we will discuss the possibility that the ALP radiation becomes non-relativistic after recombination and decays into two photons.
The photon energy density from the decay is obtained from the Boltzmann equation (see, e.g.~\cite{Jaeckel:2021gah})
\begin{align}
\non
 \dot{\rho}_{\g,k}-& Hk \frac{\partial {\rho_{\g,k}}}{\partial k}+4H \rho_{\g,k}\\
 &=\int_{-\infty}^{\infty}{d \log k' P[k, k'] \Gamma_{a}[k'] \rho_{a, k'}},\laq{chiboltz}
\end{align}
where $\rho_{\g,k}$ is defined by \eq{rhoap} (but not \eq{approx}) with $a$ replaced by $\g$ everywhere. 
The kernel,
\beq
P[k, k']= 2 (k/k')^{2}\theta{(k'-k)},
\eeq
is obtained from the phase space distribution of a two-body decay to massless modes and again we assumed $E_a \gg m_a$.  An example photon spectrum is shown as the red solid line in  Fig.~\ref{fig:1}.\footnote{We note that the spectrum is not cut off by the opacity of the Universe to photons before the recombination era. This is because the photons result from decays that happen later. 
This is in contrast to the decaying DM to photons, whose spectrum is cutoff at around $\frac{m_\phi}{2} R_{\rm rec}$. }  
We also display observed photon data, more precisely the $1\s$ upper bounds from HEAO-1~\cite{Gruber:1999yr}, INTEGRAL~\cite{Bouchet:2008rp}, COMPTEL~\cite{kappadath1998measurement}, EGRET~\cite{Strong:2004de} and FERMI~\cite{Fermi-LAT:2012edv}, which are adopted from \cite{Essig:2013goa}.
The black solid line represents  the photon spectrum from the cascade decay of the ALP originating from a decaying DM, which we will discuss in detail later (see \Sec{DM}). 

One can see that X- and $\g$-ray observations set stringent bounds on the photon spectrum from the decay of the ALP dark radiation. 
Requiring that the photon flux is below the data points, we obtain the bounds shown in Fig.~\ref{fig:1-2} in the $E_{\rm peak}-g_{a\g\g}\sqrt{\D N_{\rm eff}}$ plane. 
We do not show the region with $E_{\rm peak}< m_a$, where the approximation of relativistic ALPs is no longer valid.
The limit region also features an upper boundary. Above this the life-time of the ALP radiation is shorter than the recombination era, and the produced photons cannot freely propagate to Earth. 
 \begin{figure}[t!]
\begin{center}  
\includegraphics[width=85mm]{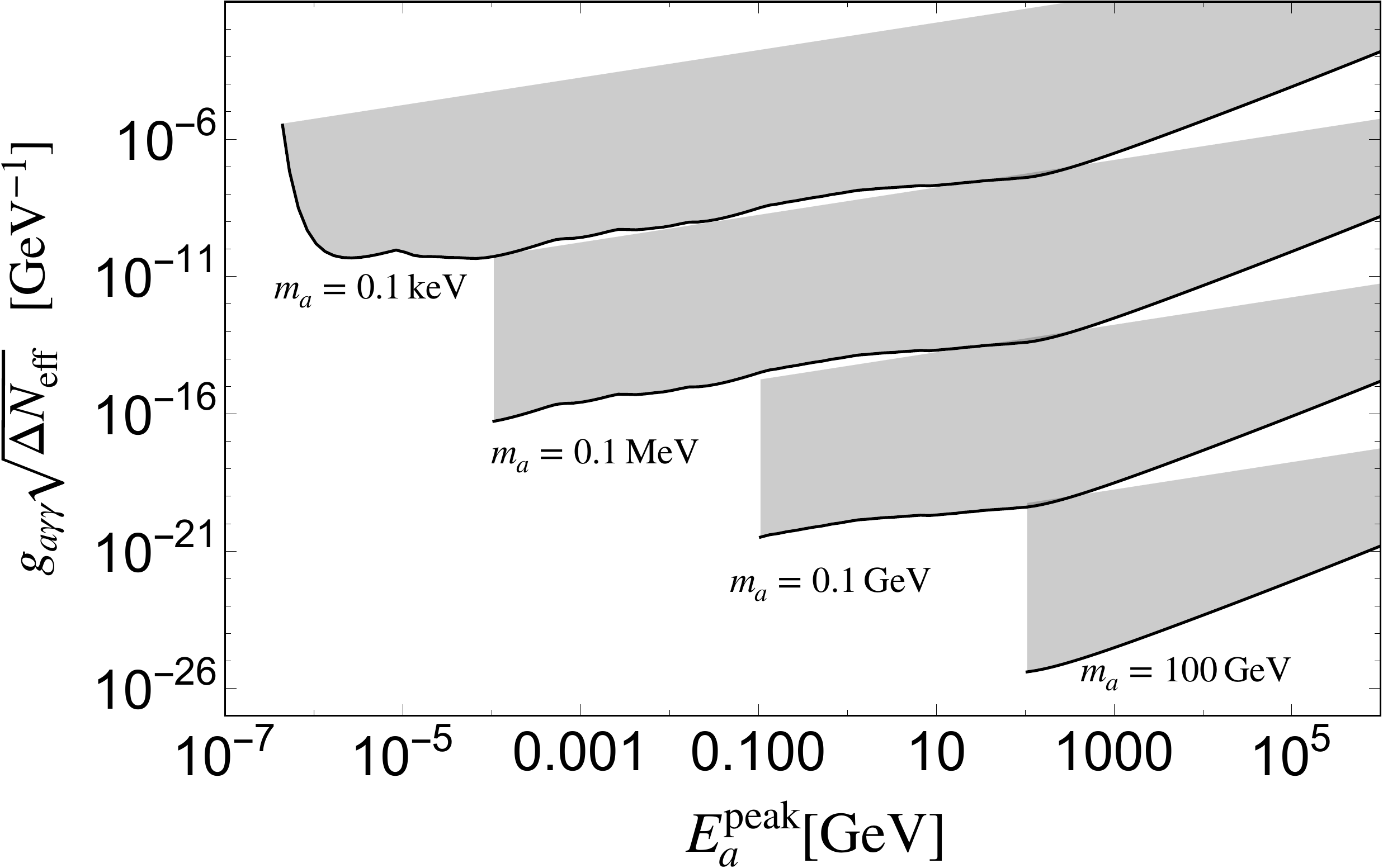}
\end{center}
\caption{The dark radiation bounds from X- and $\gamma$-ray observations~\cite{Gruber:1999yr,Bouchet:2008rp,kappadath1998measurement,Strong:2004de,Fermi-LAT:2012edv,Essig:2013goa} are shown in $E_a^{\rm peak}-g_{a\g\g} \sqrt{\D N_{\rm eff}}$ plane. From top to bottom the ALP mass is taken to be $m_a=0.1 \KEV, 0.1 \MEV, 0.1 \GEV, 0.1\TEV$. We take $\D N_{\rm eff}=0.1$. If we decrease $\D N_{\rm eff}$ the upper boundaries of the exclusions move down  by $\sqrt{\D N_{\rm eff}}$ but the lower boundaries do not change. 
}
\label{fig:1-2}
\end{figure}

 \begin{figure}[t!]
\begin{center}  
\includegraphics[width=85mm]{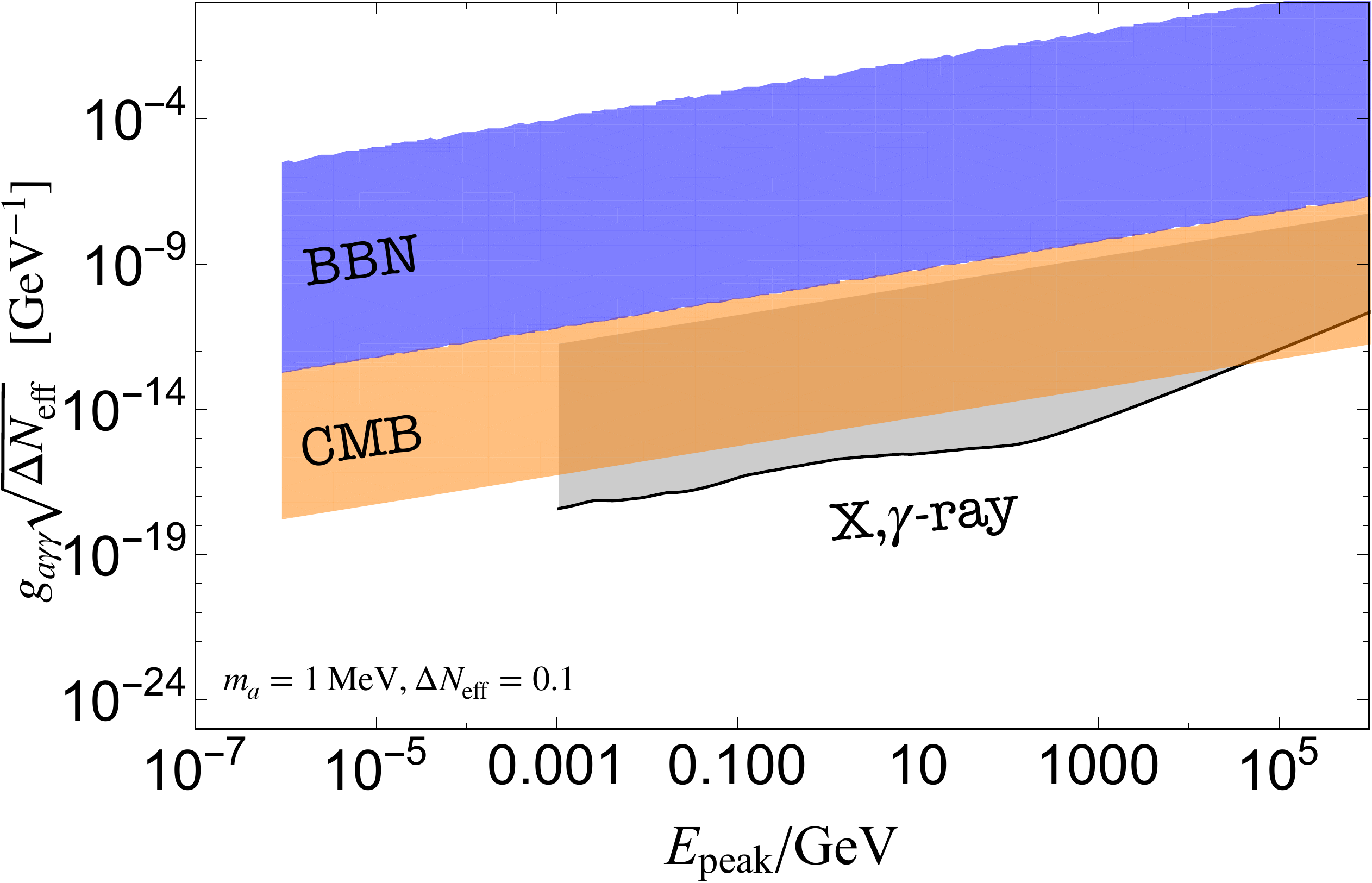}
\end{center}
\caption{The constraints for ALP dark radiation with $m_a=1\MEV,$ in the peak energy, $E_{\rm peak}$ and $g_{a\g\g} \sqrt{\D N_{\rm eff}}$ plane. 
The blue, orange, and gray regions are excluded by the bounds from BBN, CMB and X-,$\g$-ray observations, respectively, from top to bottom. 
Here we fix $\D N_{\rm eff}=0.1$ but the lower boundaries of the constraints do not change by varying  $\D N_{\rm eff}=\O(10^{-3}-1).$ 
We have adapted the BBN and CMB bounds from Ref.\,\cite{Kawasaki:2017bqm} and Ref.\,\cite{Poulin:2016anj}, respectively. As above the X-, $\gamma$-ray limits are from~\cite{Gruber:1999yr,Bouchet:2008rp,kappadath1998measurement,Strong:2004de,Fermi-LAT:2012edv,Essig:2013goa}.
}
\label{fig:2}
\end{figure}

 \begin{figure}[t!]
\begin{center}  
\includegraphics[width=85mm]{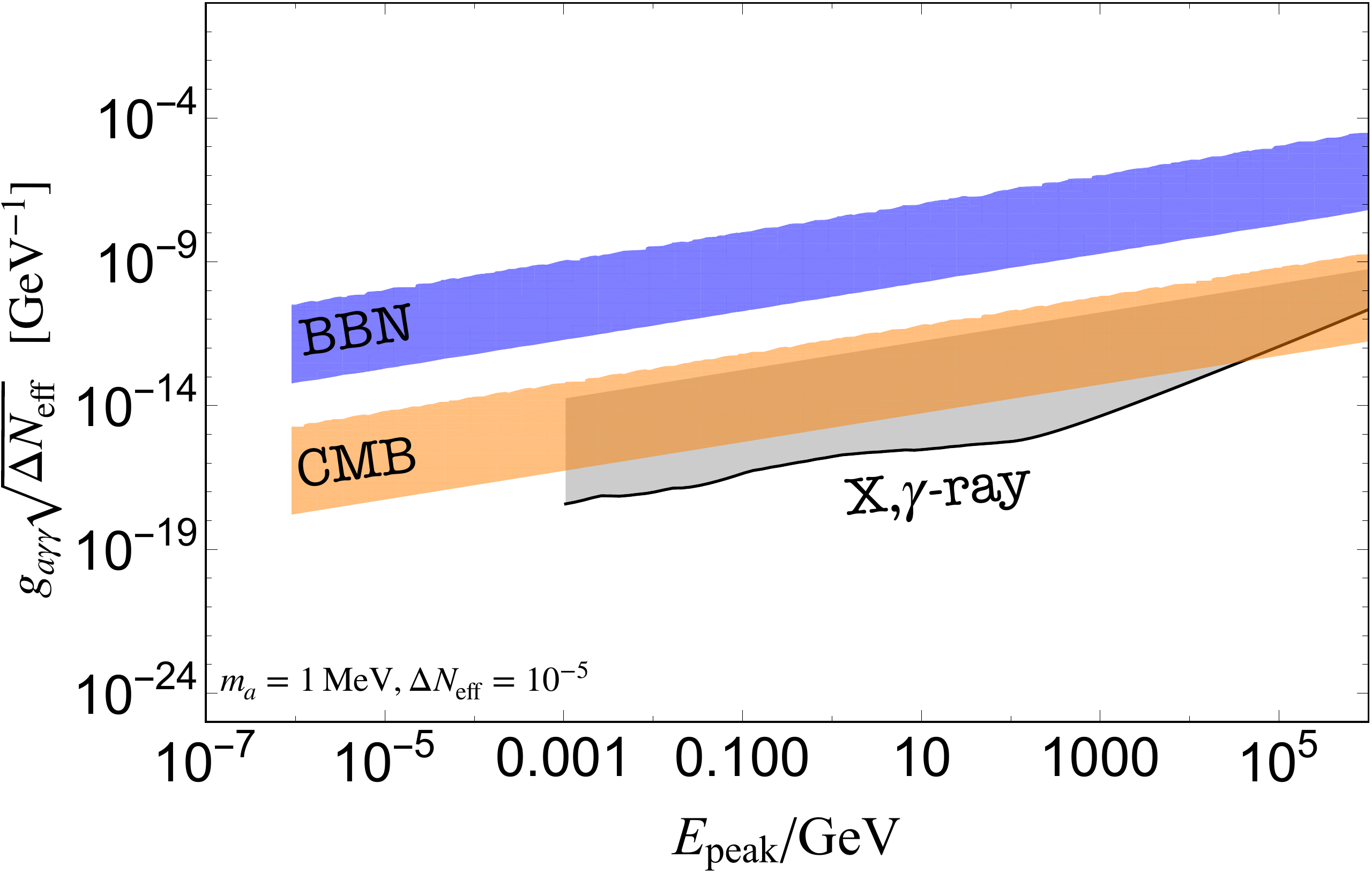}
\end{center}
\caption{Same as Fig.\,\ref{fig:2} but $\D N_{\rm eff}$ is taken to be $10^{-5}$ from which on the CMB and BBN bound starts to change, significantly. 
}
\label{fig:3}
\end{figure}

Photons from decays before recombination, on the other hand, are constrained 
by limits on the energy injection that would lead to distortions in the CMB and changes in BBN. 
A detailed simulation of this is beyond our scope. 
Here, we will instead recast the energy injection constraints from the CMB~\cite{Ellis:1990nb, Hu:1993gc, Adams:1998nr, Chen:2003gz, Poulin:2016anj,Bringmann:2018jpr, Acharya:2019uba} and BBN~\cite{Kawasaki:1999na, Kawasaki:2000en,Hannestad:2004px, Ichikawa:2006vm, DeBernardis:2008zz, deSalas:2015glj, Kawasaki:2017bqm, Hufnagel:2018bjp, Forestell:2018txr, Hasegawa:2019jsa, Kawasaki:2020qxm,Depta:2020zbh} for the decay  of non-relativistic heavy particles, which is well studied. 
To this end, let us compare the photon energy from decaying ALP DM and decaying ALP radiation.  
In both cases of a non-relativistic and a relativistic $a$ decay, the total energy deposition around the time $t$ is given by integrating the (logarithmic) energy spectrum of the ALPs multiplied by the fraction $\Gamma_{a} dt$ that decays during a given time interval $dt$. This results in, 
\begin{align}
\laq{drhog}
{\D \rho_{\g}}\propto dt \int \rho_{a, k} \G_{a}[k]  d \log{k} \\
\sim { \rho_a\frac{ \G_{a}[E_{\rm peak}]}{H} d\log{(1+z)}}.
\end{align}
In the second line we have roughly approximated the momentum integral by assuming that the (average) decay rate is given by the value at the peak momentum of the spectrum for illustrative purposes. Moreover we have replaced the time integral by one over red-shift.

Neglecting changes in the comoving ALP number, i.e. assuming that only a small fraction of ALPs decay, we have in both cases 
\beq
\G_a \rho_a \propto R^{-3} 
\eeq 
since the dark radiation  (matter) has $\rho_a \propto R^{-4}(R^{-3})$ and $\G_a\propto R (R^0)$, respectively. This scaling also holds for the ALP radiation in \Eq{drhog} before the approximation.
Therefore, as long as the comoving particle number does not change, we can reuse the known results for the non-relativistic case to estimate the energy deposition at $t$. Indeed, we can use the same formulas by replacing the ``life-time" and the energy density at any fixed cosmic time, $\hat{t}$, via
\beq
\tau_a^{\rm non-rela}\to \tau^{\rm rela}_a=1/\G_{a}[E_{\rm peak}[\hat t]] 
\eeq
and 
\beq
\rho_a^{\rm non-rela}\to \hat{\rho}^{\rm rela}_a[E_{\rm peak}[\hat t]].
\eeq
The comoving particle number is approximately constant if $t \lesssim \tau_a$, where we have defined  the decay time $\tau_a[E_{\rm peak}]$ via,
\beq
\G_a[E_{\rm peak}]  = 4H[\tau_a].
\eeq
In our analysis we take $\hat{t}=\tau_a$ to recast the bound given in Ref.\,\cite{Kawasaki:2017bqm} and Ref.\,\cite{Poulin:2016anj} for BBN and CMB, respectively. 
On the other hand, when the (exponential) decrease of the comoving ALP number cannot be neglected at $t$, the energy deposit until $t$ is different between relativistic and non-relativistic cases. 
Therefore, if $\tau_a$ is around  or smaller than the typical time-scale for the recombination or BBN era, the energy deposits are different and our analysis is not very accurate. 
On the other hand, if $\tau_a$ is much larger than those scales, our analysis should be valid.

The X- and $\g$-ray bound (gray region) as well as the recast CMB (orange region) and BBN (blue region) bounds are shown in Fig.~\ref{fig:2} and \ref{fig:3} in the $E_{\rm peak}$ and $g_{a\g\g} \sqrt{\D N_{\rm eff}}$ plane. 
In Fig.~\ref{fig:2}, we fix $\D N_{\rm eff}=0.1$. That said, as long as $\D N_{\rm eff}= \O(10^{-3}-1)$, the limits are very similar. 
We do not show the X- and $\g$-ray bounds for the ALP energy {beyond} $E_{\rm peak}= m_a$ below which the relativistic approximation for the ALP at the present Universe is no longer valid.  
The cut at the upper energy of the BBN and CMB bounds are taken to be $E_{\rm peak} R_{\rm rec}= m_a$ where $R_{\rm rec}$ is the scale factor at recombination. The lower boundary of $g_{a \g\g}$ should be accurate, 
but the upper boundary of the BBN one may be considered as an order of magnitude estimate {as we have discussed previously}.

In Figs.~\ref{fig:41}, \ref{fig:42}, \ref{fig:43}, and \ref{fig:44} the bounds are represented in the $m_a-\sqrt{N_{\rm eff}}g_{a\gamma \g}$ plane by fixing $p_a^{\rm peak}=10\KEV, 1\MEV, 1\GEV, 0.1\KEV$, respectively. 
For comparison we also show the constraint from the extra cooling of Horizontal branch stars as the purple dashed line (adopted from Refs.\,\cite{Ayala:2014pea,Carenza:2020zil}; see also Refs.\,\cite{Raffelt:1985nk,Raffelt:1987yu,Raffelt:1996wa}). This is the only bound strongly sensitive to $\D N_{\rm eff}$. In the figures we take  $\D N_{\rm eff}=0.1$.
In addition we also indicate the photon constraints for the situation when the decaying ALP is non-relativistic at present but it is relativistic at the recombination era, i.e. $ p_a^{\rm peak}\lesssim m_a \lesssim p_a^{\rm peak}/R_{\rm rec}$. This bound is obtained by the following simple procedure. 
We assume that the ALP contributes to today's cold DM with density,
\begin{equation}
    \rho_a^{\rm nr}=\frac{m_a}{p_a^{\rm peak}} \D N_{\rm eff} \frac{7}{4}\frac{\pi^2}{30} T_{\rm \nu}^4
\end{equation} 
with $T_\nu$ being the present neutrino temperature.

Then we can estimate the photon flux from the non-relativistic ALP decays by {the formula (see~\Eq{approx} and the comments below it)}
\beq 
E^{2}_{\g}\frac{d^2 \F_\gamma }{d E d \Omega} = 16 E_\g^4 \frac{\Gamma_{a\to \g \g}[p_a=0] \rho_{a}^{\rm nr}}{4\pi H(t') R[t']^3 m_a^4}
\eeq 
in the energy range of 
\beq 
p_a^{\rm peak}/2 \leq E_\gamma \leq m_a/2.
\eeq 
This photon spectrum looks like the black dotted line in Fig.~\ref{fig:1} without the line peak.
Being a bit more precise, the part of the spectrum $E_\gamma< p_a^{\rm peak}/2$ mostly comes from the era when the ALP  peak momentum $p^{\rm peak}_a /R[t'] >m_a$ and thus the ALP becomes relativistic and our estimate is invalid. The spectrum there, however, is suppressed and we do not consider it.  We can compare the spectrum with the data points of X- and $\gamma$-ray to display the dotted lines in Figs.~\ref{fig:41}, \ref{fig:42}, \ref{fig:43} and \ref{fig:44}. 
This bound as well as the CMB and BBN bounds\footnote{$p_a^{\rm peak}/R_{\rm rec}>$ several keVs is needed to apply {and recast the} CMB bound.} can constrain the dark radiation (or the redshifted non-relativistic ALP) even if 
the peak momentum is smaller than the mass. In particular they are important when the peak momentum is much smaller than {a} keV (e.g. Fig.~\ref{fig:44}), in which case we cannot use the limits from the relativistic ALP decays.

Our analysis for the non-relativistic regime of the ALPs should only be taken as an order of magnitude estimate when $p_{a}^{\rm peak}/R_{\rm rec} \lesssim 2m_{a}$ in which case the ALP velocity today is smaller than the escape velocity of the Milky-way galaxy. Then, the spatial distribution may be non-trivial.\footnote{In the case that the ALP is lighter  than the inequality by a factor of a few, we may have an observational anisotropy of the X-, $\gamma$-rays in particular in the direction of galaxy clusters, which have larger escape velocity than the Milky-way galaxy.} We expect when $p_{a}^{\rm peak}/R_{\rm rec} \gg 2m_{a}$, i.e. in most of the range shown in the figures, the derived constraint is reasonably accurate.
Throughout the numerical analysis in this paper, we have also neglected the effect of the ALP number change due to its decays, which should modify the resulting photon spectrum when $\D N_{\rm eff}$ is small and $m_a$ is large (see e.g. Ref.~\cite{Ho:2019ayl}.).

We also note that $\rho_a^{\rm nr}/\rho_{\rm DM}\simeq 0.05 \times \Delta N_{\rm eff} \frac{\MEV}{p^{\rm peak}_a} \frac{m_a}{1\GEV} $, which is smaller than $\O(1)\%$ in the whole parameter region shown in the figures. This implies that the bounds from limits on hot DM are not severe.
  
In any case, when $m_a \gtrsim \KEV$ and $\D N_{\rm eff}$ is not extremely suppressed, the bounds on ALP dark radiation from X-/$\gamma$-ray, CMB and BBN can be more stringent {than other} existing bounds. 

\begin{figure}[t!]
\begin{center}  
\includegraphics[width=85mm]{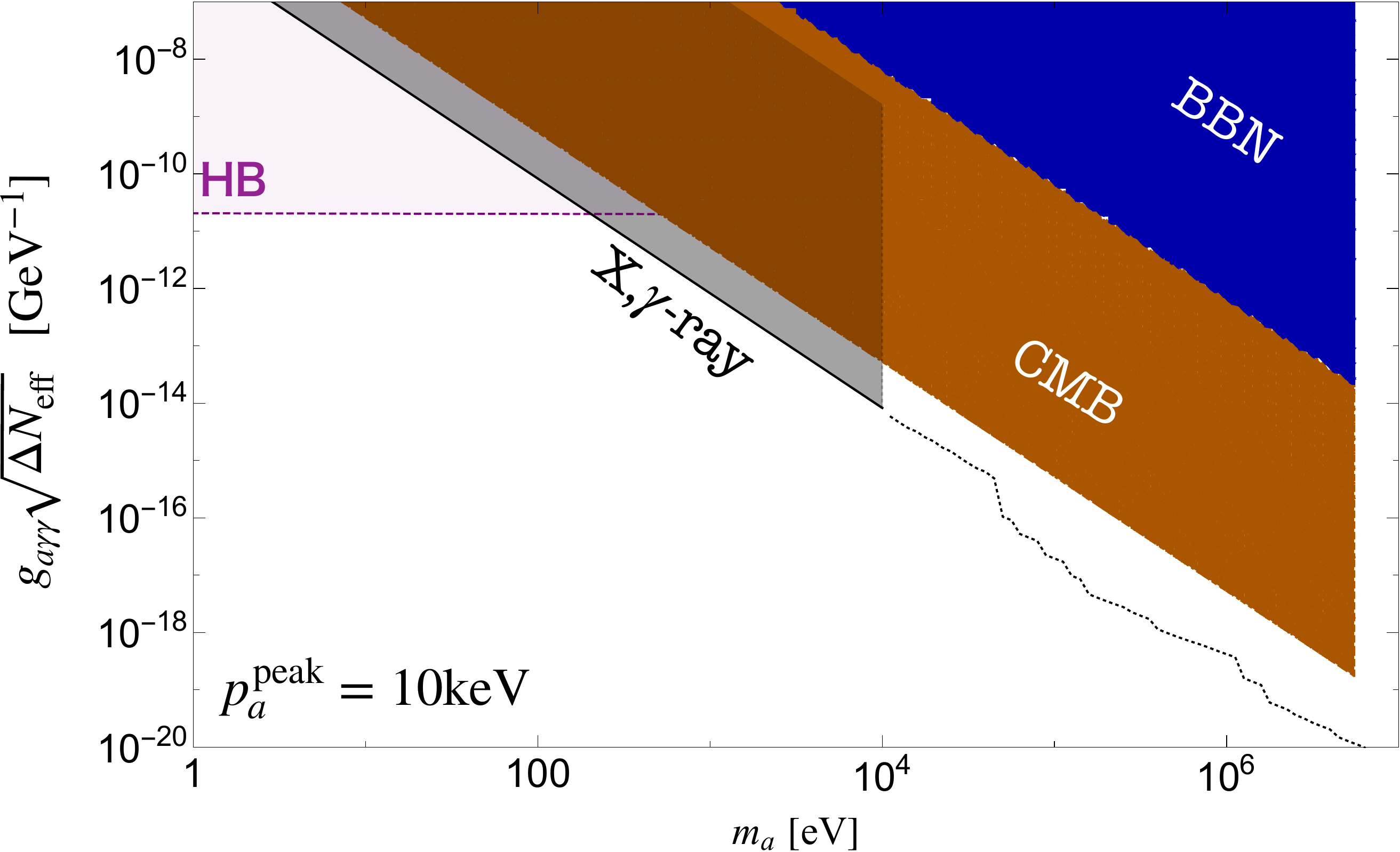}
\end{center}
\caption{Same constraints as Fig. \ref{fig:2} but in the $m_a-\sqrt{\D N_{\rm eff}}g_{a\gamma\gamma}$  plane and fixing $p_a^{\rm peak}=10\KEV$.  $\D N_{\rm eff}=0.1$ is taken. The cooling constraint from Horizontal branch stars~\cite{Raffelt:1985nk,Raffelt:1987yu,Raffelt:1996wa,Ayala:2014pea,Carenza:2020zil} is shown by the purple dashed line adopted from Ref.\,\cite{Ayala:2014pea,Carenza:2020zil}.
The constraints from photons when the ALP is non-relativistic today is shown by the black dotted line by taking only the extra-galactic component into account.
By decreasing $\D N_{\rm eff}$ the purple dashed line moves downwards, but {other lower boundaries} of the bounds do not change much for $\D N_{\rm eff}>10^{-3}.$}

\label{fig:41}
\end{figure} 
 
  \begin{figure}[t!]
\begin{center}  
\includegraphics[width=85mm]{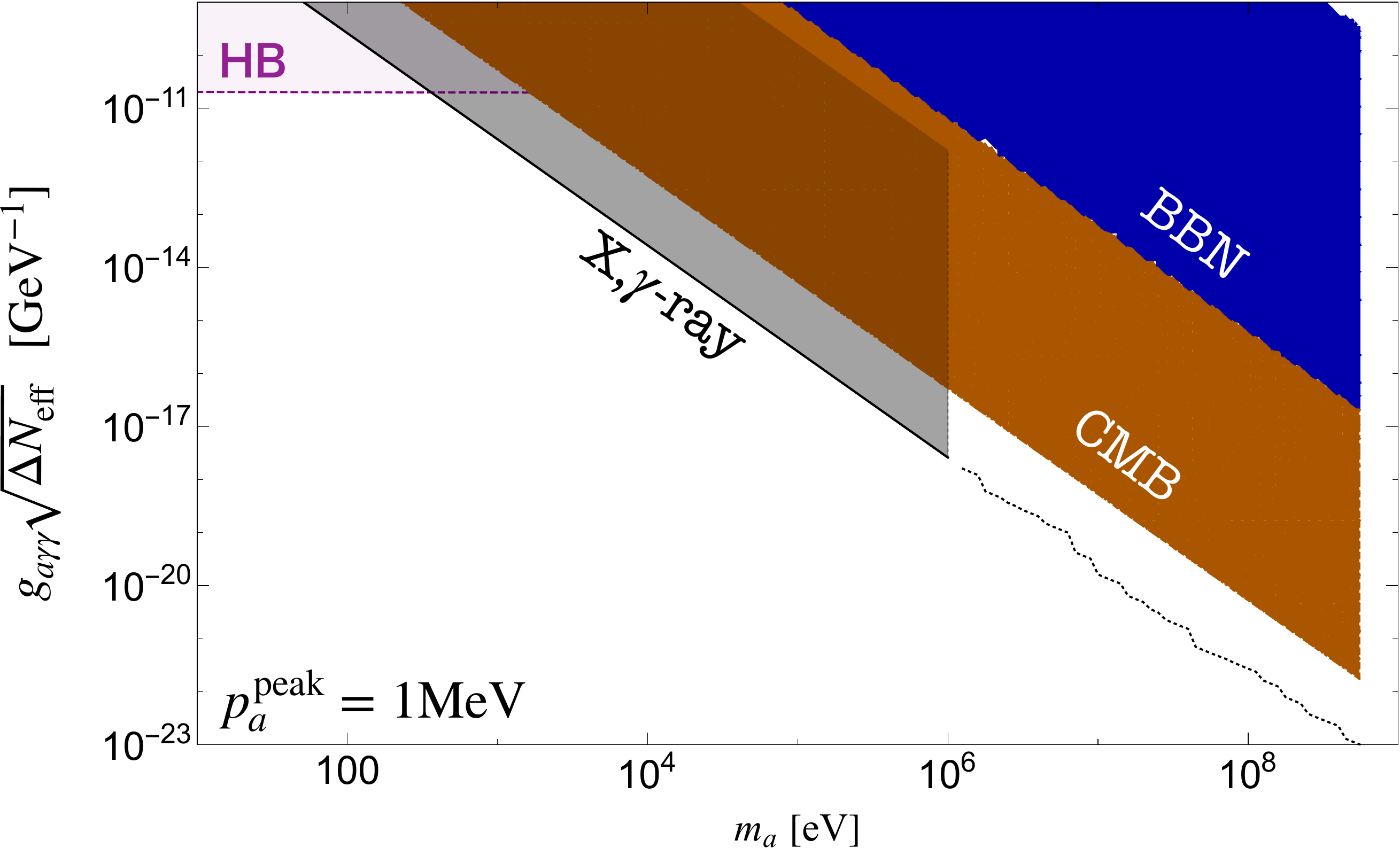}
\end{center}
\caption{Same as Fig.\,\ref{fig:41} with $p_a^{\rm peak}=1\MEV$. 
}
\label{fig:42}
\end{figure} 
 
  \begin{figure}[t!]
\begin{center}  
\includegraphics[width=85mm]{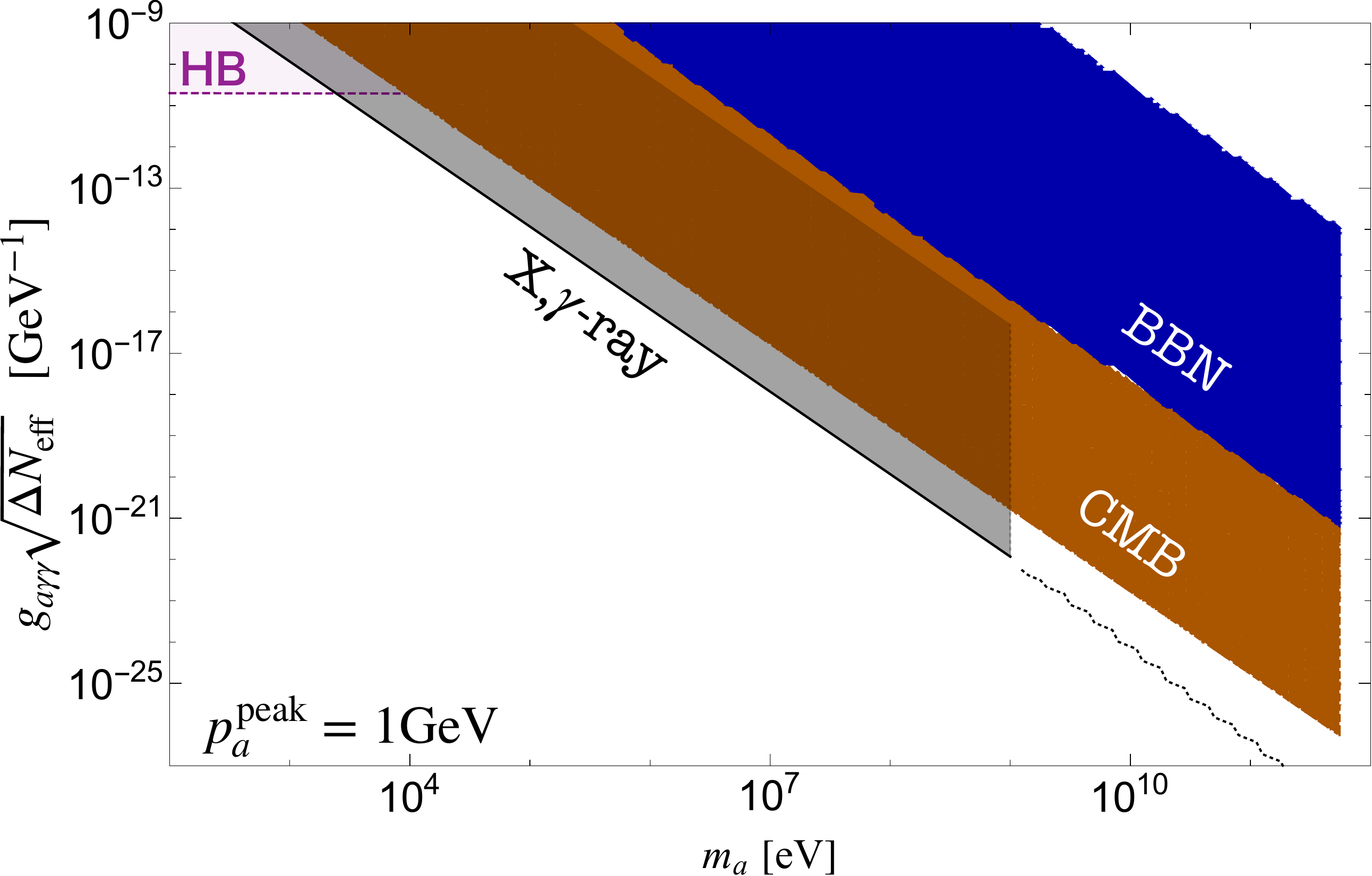}
\end{center}
\caption{Same as Fig.\,\ref{fig:41} with $p_a^{\rm peak}=1\GEV$.
}
\label{fig:43}
\end{figure} 
 
   \begin{figure}[t!]
\begin{center}  
\includegraphics[width=85mm]{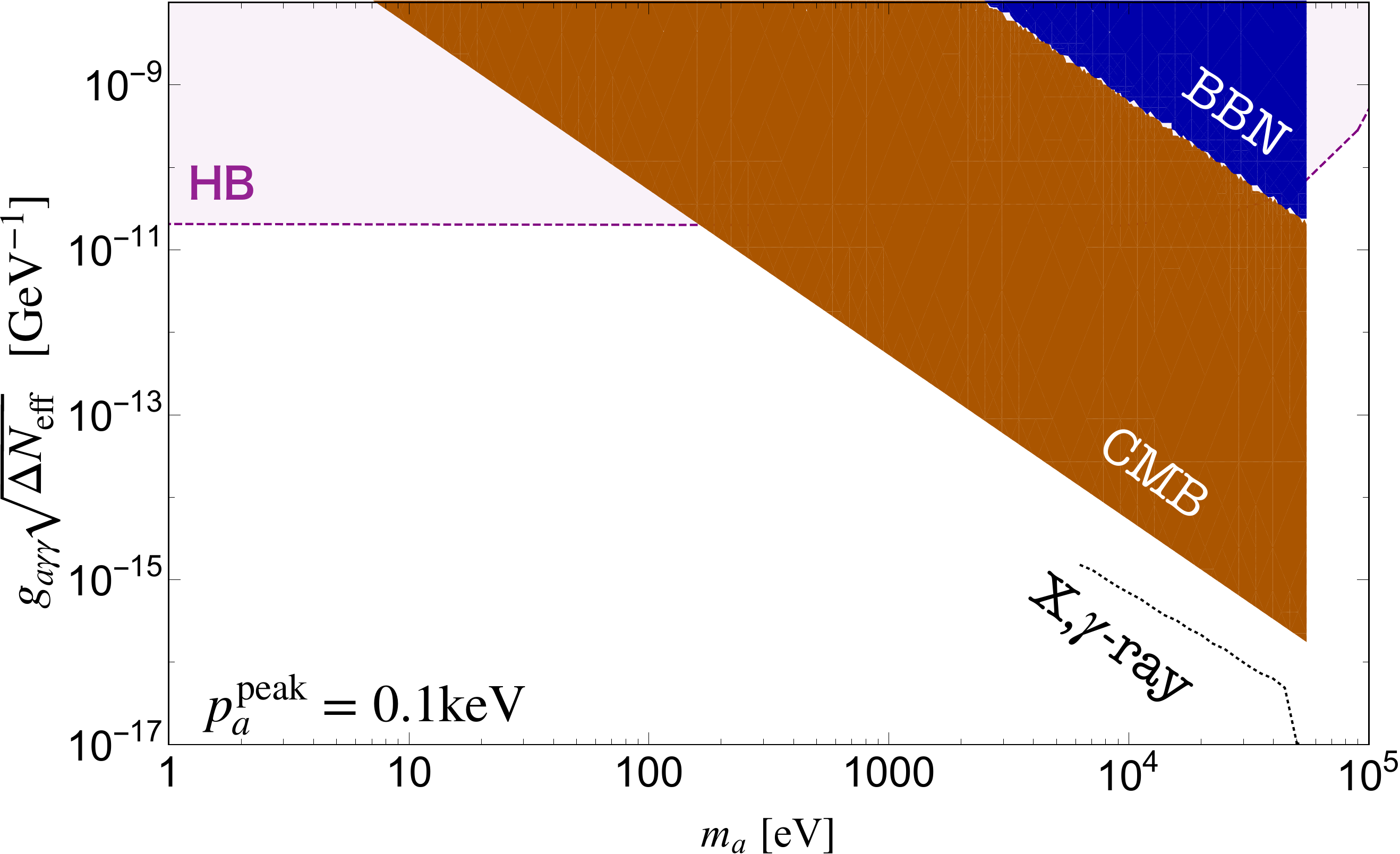}
\end{center}
\caption{Same as Fig.\,\ref{fig:41} with $p_a^{\rm peak}=0.1\,{\rm keV}$. Since the peak energy is low the X-, $\gamma$-ray constraint are only efficient when the ALP is heavy and becomes non-relativistic until the present.
}
\label{fig:44}
\end{figure}

As mentioned, our constraint does not change much if the ALP radiation is produced from a decaying mother particle that does not reheat the Universe. 
In the case of a two-body decay, the spectra of the ALP radiation only differs by a little~\cite{Conlon:2013isa, Jaeckel:2021gah}. 
However the tiny difference of the ALP spectrum implies that we can possibly measure the reheating through it~\cite{Jaeckel:2021gah}.

\section{Cascade photons from DM decaying to ALPs}
\lac{DM}

So far we have studied the situation where the dark radiation ALPs originate from the decay of a non-relativistic modulus in the very early Universe. However, if the decay happens much later and does not complete until the present
the spectrum of $a$ will be very different. 

Let us assume that the mother particle $\f$ becomes non-relativistic much before  matter-radiation equality, but decays after this time.\footnote{A scenario in this spirit with DM particles decaying into axions (plus photons) has, e.g. recently been discussed in~\cite{Bae:2018mgq,Gu:2021lni}, albeit for lower axion/ALP mass.
} Then their spatial distribution is affected by structure formation, as they essentially behave like a fraction of the DM. In particular, $\phi$ will also be more concentrated in regions of high DM density such as the galactic center.
In this situation it is clear that we have a non-trivial spatial distribution. However, due to red-shifting galactic and extra-galactic components will also have different spectra.

We start by considering the galactic and extra galactic components of the ALP spectrum and then discuss the corresponding photon spectra,
\beq
\rho_{a,E_a}=   \rho_{a,E_a}^{\text{extra}}+\rho_{a,E_a}^{\rm galactic}.
\eeq
As already noted the angular distributions of  the components are 
different.
Clearly the ALPs from the galactic component mostly come from the galactic center.

\subsubsection{Extra galactic component}
The discussion of the previous sections can be straightforwardly applied to the extra-galactic component from \Eq{rhoap} with $\rho_\phi \propto R^{-3},$ i.e. 
\beq
\rho_{a ,E_a}\propto E_a \frac{\rho_c \Omega_\f  \G_{\f\to aa }  \Theta[m_\f/2-E_a]  }{H|_{z=m_\f/2E_a -1}}
\eeq 
where $\rho_c$ is the critical density, $\Omega_\f$ is the abundance of $\f$ today, and we neglect the decrease {in} the comoving number of $\f$. 
In the dark energy dominated Universe $\rho_{a,E_a}\propto E_a$ and 
it has a sharp cutoff at $E_a=m_\f/2.$
We can solve \eq{chiboltz} to obtain the photon spectrum due to the ALP decay. 
The phase space suppression leads to the decrease of the spectrum with $\rho_{\g,E_\g }\propto |m_\f/2 -E_\g |$ when $E_\g \sim m_\f/2$. 
On the other hand, when $E_\g \ll m_\f/2$ the spectrum is similar to the moduli decay case at lower energy (neglecting the difference of $z$ dependence in $H$). 
Around the energy where the phase space is suppressed, the effect of the red-shift is neglected. 
Then the photon energy is peaked at around $m_\f/4$ since it is from the decay of  ALP with $E_a=m_\f/2$. 
The behavior agrees well with the numerical result presented  in Fig.~\ref{fig:1} with $\Omega_\f h^2=0.12. $
The extra galactic component is dominant compared to the galactic component if the ALP decay life time is not  too short as we will see next.

\subsubsection{Galactic component}

Due to the gravitational interaction a fraction of the non-relativistic $\f$ gathers around the galactic center. 
Since the distance from the galactic center is short, the red-shift of the energies can be neglected. 
Then the ALP flux has an energy $m_\f/2,$ which is represented by 
\begin{align}
\label{galacticALP}
\frac{d\Phi^{\rm galactic}_a}{dE_a }
&=\int ds d \Omega \frac{\Gamma_{\f\to aa} }{4\pi s^2}  \left(\frac{\rho_\phi(s, \Omega )}{m_\phi}\right) \, s^2\, \frac{d N_a}{dE} 
\end{align}
where $s$ is the direction from {the} Sun along the line of sight, and $\Omega$ is the angular direction. Moreover, we have,  
\beq
\frac{d N_a}{dE} = 2 \delta\left(E - \frac{m_\phi}{2}\right). 
\eeq
We can now use the D-factor usually defined for decaying DM (cf., e.g.~\cite{Combet:2012tt}),
\beq
D[\Omega]\equiv \int_{\Delta \Omega} d \Omega d s \rho_{\rm DM} (s , \Omega ) ,
\eeq
where $\Delta \Omega$ is the spatial angle covered by the object/region we are looking at.
Combining this with the assumption, 
\beq
\rho_\phi(s, \Omega) \approx \rho_{\rm DM}(s,\Omega) \frac{\Omega_\phi}{\Omega_{\rm DM}},
\eeq
i.e. $\rho_\f$ has a  distribution similar to that of DM, $\rho_{\rm DM}$, we can then calculate the ALP flux in a given direction,
\begin{equation}
\laq{galacticALPdiff}
\frac{d^2\Phi^{\rm galactic}_a}{dE_a d\Omega}
=\frac{D(\Omega)}{\Delta \Omega} \frac{\Gamma_{\f\to aa} }{(2\pi)m_{\phi}}  \delta\left(E - \frac{m_\phi}{2}\right) .
\end{equation}
Depending on the angular direction, the ALP flux from the decaying DM may have a dominant contribution of galactic origin. The resulting ALP spectrum is same as the photon spectrum of the dashed line 
shown in Fig.~\ref{fig:1}, where the peak represents the ALP flux from the direction towards the center of galaxy. 
We take $m_\f=1\GEV$ and $\G_\f=10^{-10}H_0$. We assume the EinastoB distribution given in \cite{Cirelli:2010xx} for illustrative purpose.\footnote{We have estimated the height of the monochromatic peak by assuming a energy bin with width of $\D E/E=0.01$.} One can see that the peak height is larger than that of the extra-galactic component. 

\bigskip 

So far this is identical to the case of an ordinary two-body decay of DM, e.g. into photons. However, up to now we have only considered the ALP spectrum. But our observable are the photons from a subsequent decay of the ALPs.
As we will see momentarily this changes the situation significantly. 

The photons from the decay of a relativistic ALP of energy $E$ have the distribution of $d N_\g^{(a, E)}/d E_\g \simeq  2/E \Theta[E-E_\g]$ if we neglect the mass of $a$ in the distribution. 
By noting that the photon is produced almost along the line of $a$-motion, 
the photon flux is approximated by
\begin{align}
\frac{d\Phi^{\rm galactic}_\g}{dE_\g }\approx &\int_0^\infty ds  d\Omega dE  \frac{d N_\g^{(a, E)}}{dE_\g}   (1-\exp{(-s \cdot \G_{a\to \g\g})})  \non\\ 
&\times \frac{\Gamma_{\f\to aa} }{4\pi s^2}  \left(\frac{\rho_\phi(s, \Omega )}{m_\phi}\right) \,s^2 \, \frac{d N_a}{dE}\\
\approx  8\G_{a \to \g\g} & \frac{\Gamma_{\f \to aa}}{4\pi m_\phi^2} \Theta{[m_\f/2-E_\g]}
\int_0^\infty ds d\Omega
{ s}  \rho_\phi(s, \Omega ),
\end{align}
where in the last line we have assumed that the decay rate is small compared to the distance.
Compared to Eq.~\eqref{galacticALP} we have an additional factor of $s$ in the integrand.

If a DM structure is localized around a given distance $d$ from us, we can approximate 
\begin{align}
\int_{\D \Omega} ds d\Omega
 s  {\rho_\phi(s, \Omega )}\sim d \times D[\Omega ]  .
\end{align}
Therefore this component is important when the $D$ factor and the distance $d$ is large. 

By integrating the EinastoB distribution for  Milky Way Galaxy, the energy flux is peaked at $\sim 5\times 10^{-6} \MEV {\rm cm}^{-2}{\rm s}^{-1} {\rm sr}^{-1}$ with the parameter set for the cascade-decaying DM used in Fig.~\ref{fig:1} (black solid line). 
We take $\Gamma_{\phi \to aa}= 2\times 10^{-2}H_0, g_{a\g\g}= 10^{-18}\GEV^{-1}, m_\f=1\GEV, m_a=\MEV.$ 
Therefore the flux averaged by the angular integral is much smaller than the extra galactic component as can be expected from the suppression of the decay volume $ r_{\odot} H_0.$ 

ALPs from far away galaxies can decay more efficiently due to the longer decay volume. Let us consider an $\O(100)$ Mpc distant cluster, e.g. the Ophiuchus galaxy cluster. We can then estimate~\cite{Combet:2012tt},
\begin{equation}
E_\g^2 \frac{d^2\F_\g^{\rm galactic}}{d E_\g d \Omega  }\sim  \O(10^{-4}) \frac{\MEV}{ {\rm cm}^2\,{\rm  s} \,{\rm sr}}\frac{E^2_\g}{\GEV^2 } \Theta(0.5\GEV -E_\g),
\end{equation}
for the parameter set of the black solid line in Fig.~\ref{fig:1}. 
Although the galactic component of the ALP flux is comparable or a factor of a few larger than the extra-galactic one, 
the photon flux is dominated by the extra-galactic one, which is two orders of magnitude larger than the galactic component. 

\subsubsection{Discussion}

Since the spectrum from the extra-galactic component is similar to that of the moduli decay, 
we expect the lower bound  from X-rays and $\g$-rays to be similar, with $\D N_{\rm eff}$ estimated by $\rho^{\rm extra}_{a,E}[z=\O(1)].$ We emphasize again that, neglecting the galactic component in estimating the X-ray and $\g$-ray constraint, is justified when the ALP radiation has a life-time longer than the age of the Universe. 
When the life-time of $a$ is much shorter than the distance to the galactic center,  the contribution from the galactic component can be more important than the extra galactic case.\footnote{A generic cascade decay of DM with a suitable decay length for the ALP may be useful to explain non-trivial anisotropies in the distribution of cosmic-rays a la ultra-high energy cosmic rays above EeV e.g.~\cite{TelescopeArray:2014tsd} (see, however,~\cite{TelescopeArray:2018qwc}), or something like the 3.5~keV line~\cite{Bulbul:2014sua,Urban:2014yda,Boyarsky:2014ska,Cappelluti:2017ywp,Dessert:2018qih}.
For example, if they are disfavored to originate from the galactic center of the Milky-way,
a suitable decay length may lead to a relative brightening of objects at a distance of the order of the decay length.}
Therefore the X-ray and $\g$-ray bound becomes even more stringent if the life time of ALP radiation is shorter than the time scale of recombination. This is different from the case of ALP radiation from the early Universe, which was our main topic.

Lastly, let us mention the possibility to distinguish between an early and a late decaying
precursor particle, $\f$, from which the ALP is produced. 
When $\f$ is the DM and decays late, the spectrum of the resulting photons is different from the early decay case and thus we can distinguish it if we have enough energy resolution (e.g. there are various experiments with energy resolution smaller or even much smaller than $\O(0.1)$ \cite{Barret:2018qft,CTAConsortium:2010umy,eROSITA:2012lfj,Fermi-LAT:2012edv,Fermiweb,Galper:2012fp,XRISMScienceTeam:2020rvx}.)
Another possible way to distinguish the two options is the angular distribution, which is more pronounced for the 
late decay case. While, as we have seen, individual structures are not exceptionally bright if the decay length is large, it may nevertheless be possible to pick them out, in particular since they feature a slightly different energy spectrum. For example, let us look at the galactic center. Decay photons from there are not affected by red-shift and the peak of their spectrum is at $\sim m_\f/4$. Therefore, this region of the spectrum should be slightly brighter when looking in the direction of the galactic center.
This feature is absent if the ALP is produced from early moduli decay. 
In addition, in the case of late decay,
$\D N_{\rm eff}$ measured by the CMB data is smaller than that for the decay to radiation around today, while the decaying dark radiation has same $\D N_{\rm eff}$ (or larger if the dark radiation mostly decays to the photon).

\section{Conclusions}
In this paper, we have calculated limits on ALP dark radiation, made from massive ALPs decaying into photons, from the observation of X- and $\g$-rays but also from recasting constraints on the energy injection during the CMB and BBN eras. 
We have focused on the case where the ALPs are produced in the decay of a heavy non-relativistic precursor that could, e.g. be the inflaton or a modulus. In such, often string inspired, scenarios the reheating temperature is {typically} relatively low, thus our limits complement those on thermally produced heavy ALPs~\cite{Cadamuro:2011ti,Cadamuro:2011fd}.

It should be mentioned that our bounds depend on the peak energy of the ALP radiation spectrum. As discussed, this is model depend and hence is, perhaps the main loophole to our analysis. 

Keeping the just mentioned caveat in mind, if the mass is above MeV and $\D N_{\rm eff}$ is not too small, we can constrain photon couplings of the ALP originating from scales as high as the string or even Planck scale. 
In this sense, future observations of X- and $\g$-rays as well as the CMB provide an interesting opportunity to test ALPs with potentially stringy origin.

\section*{Acknowledgments}
JJ would like to thank A. Hebecker and M. Wittner for discussions and collaboration on stringy setups with ALP dark radiation.
WY was supported by JSPS KAKENHI Grant Number 20H05851 and 21K20364 \\

\bibliography{references}

\end{document}